\documentclass[aps,pra,twocolumn,preprintnumbers,amsmath,amssymb,floatfix]{revtex4-1}

\usepackage{graphicx}
\usepackage{dcolumn}
\usepackage{epsfig}

\begin{document}

\preprint{APS/123-QED}

\title{Empirical relationships for heavy-ion equilibrated charges and charge-changing \\ cross-sections in rarefied hydrogen and their application}

\author{R.N. Sagaidak}
 \email{sagaidak@jinr.ru}
\affiliation{Flerov Laboratory of Nuclear Reactions, Joint Institute for Nuclear Research, J.-Curie, 6, 141980 Dubna, Moscow region, Russia}

\date{\today}

\begin{abstract}
Ionized heavy evaporation residues (ERs) resulting from heavy ion (HI) fusion-evaporation reactions are knocked out from solid targets to rarefied gas of gas-filled recoil separators. In gas, ionized ERs undergo charge-changing collisions on their way to a detection system. An equilibrium between the loss of charge (electron capture) and the gain of charge (electron loss) for ionized ERs allows one to use equilibrated charge state distributions in trajectory calculations for ERs moving through a magnetic field. The distance from a target to the point where the ER charge state distributions become to be the equilibrated ones is essential for such calculations. This distance can be estimated in simulations based on electron capture and loss cross-sections for energetic HIs. Proper approximations of available experimental data have provided these cross-sections, which were applied to the simulation of the ionic charge evolution for ionized heavy ERs.
\end{abstract}

\maketitle

\section{Introduction}
\label{intro}

Gas-filled recoil separators (GFRS) are a useful tool for the separation and studies of the heavy evaporation residues (ERs) produced in complete-fusion reactions of heavy ion (HI) projectiles with heavy target nuclei \cite{Karn69Bach70,Miyat87,Ghio88,Ninov95,Leino95,Subot02,TASCA08,SHANS13}. In these devices, ERs recoiling out of a thin target are efficiently separated in-flight from the primary HI beam particles and other unwanted products within a separation time $\sim$1 $\mu$s. A separation (collection) efficiency for ERs is essentially determined by their kinematics (angular and energy distributions) and abilities of deflecting and focusing elements used in a particular separator.

Larmor radius $\rho$ of HI with atomic mass number $A$, ionic charge $q$, and velocity $v$ in a homogeneous magnetic field with a magnetic flux density $B$ is defined by magnetic rigidity. Magnetic rigidity $B\rho$ (in Tm) is given by the following relation:
\begin{equation}
\label{Brho}
  B\rho = 0.0227 A (v/v_{0}) / q = 0.144 (E A)^{1/2} / q,
\end{equation}
where $v_{0}$=2.187$\times$10$^{6}$ m/s is the Bohr velocity, and $E$ is the HI energy in MeV. In the region filled with gas, HIs undergo atomic collisions, in which electrons can be either lost or captured, changing the charge states of HIs. If the mean free path of HIs between two charge-changing collisions is short enough compared with the length of its trajectory, i.e., the frequency of charge-changing collisions is large, the charge states are well defined. In these conditions, HIs closely follow the trajectory determined by the magnetic rigidity, corresponding to their mean (equilibrated) charge value. According to the Bohr predictions \cite{Bohr41}, all electrons with orbital velocities smaller than $v$ are stripped, so mean charge $q_{m}$ of HI with atomic number $Z$, as it is derived with the Thomas-Fermi model, can be expressed by
\begin{equation}
\label{BohrCha}
    q_{m} = (v/v_{0}) Z^{1/3}
\end{equation}
for the velocity range of $1 < v/v_{0} < Z^{2/3}$. Then Eq.~(\ref{Brho}) can be rewritten as
\begin{equation}
\label{BrhoBohr}
  B\rho = 0.0227 A / Z^{1/3}.
\end{equation}
Hence corresponding trajectories are determined by the mass and atomic numbers only and are independent of the initial charge and velocity distributions.

In the present work, the emphasis is on the estimates of how far from a target, the equilibrated charge-state distribution for ERs moving in rarefied H$_{2}$ is settled. These estimates are essential for a new generation of GFRS intended for the synthesis of superheavy nuclei (SHN) in complete-fusion reactions with accelerated HI beams \cite{OgaUtNPA15}. For example, a new Dubna GFRS allows using rather thick targets ($\sim$1 mg/cm$^{2}$) located upstream a beam just in front of the first magnetic quadrupole lens of the separator \cite{Popeko16,Beeckman19}. A thick target and its proximity to the first optical element should radically increase a detected yield of SHN formed with extremely low production cross-sections \cite{OgaUtNPA15}, despite the increased spread in the energy and angle of escaped ERs.

In general, ERs escaping a target have a very broad charge state distribution. In addition to the equilibrated `solid' component (see, for example, \cite{SIM82,SY94,SG01}), there is a non-equilibrated one with charge states much higher than those inherent in the equilibrated charges \cite{Steig63,Scob83,Brink94,Saga08,Saga18}. This non-equilibrated component corresponds to excited nuclear states, which strongly affect the ionization of inner atomic shells owing to the conversion of nuclear transitions in ERs. Vacancies formed in the conversion of inner shells of ionized ERs lead to the Auger cascades, which significantly increase the ion charges of ERs over the expected equilibrium magnitudes. Thus, reliable electron capture and loss cross-section data for highly charged HIs are necessary to trace the evolution of ER charge state distributions from the initial to the equilibrated one.

In the next section, systematics of the mean equilibrated charges for HIs and ERs in H$_{2}$ are considered. They are further used for the approximation of the charge-changing cross-section data. The empirical approximations of available data on electron capture and loss cross-sections are considered in Section \ref{CSparam}. Monte Carlo simulations of the evolution of charge-state distributions for ERs are described in Section \ref{MCsim}. The results are summarized in Section \ref{Summa}.

\section{Mean equilibrated charges for heavy ions and ERs in H$_{2}$}
\label{MeanChar}

Usually, a HI charge-changing process in gas is considered in the approximation of proximity to the equilibrated charge with taking into account single-electron capture and loss cross-sections. They are the exponential functions of HI charge state $q$ and have a fixed value of $q_{m}$ as a parameter \cite{Ninov95,Betz73,Paul89,Gregor13}. Different approaches were used for the estimate of $q_{m}$. Thus in \cite{Ninov95,Paul89}, the empirical systematics based on the data available at that time \cite{WitBetz73,DmiNik64} were used. In another approach \cite{Ghio88}, the parametrization according to \cite{Betz73} and available experimental data was used to obtain the dependence of $q_{m}$ on $Z$ and $v/v_{0}$ in He gas. Later, the same data were used for new systematics \cite{Gregor13}, in which additional data were included.

\begin{figure}[!t]
\vspace{0.5mm}
\centerline{\includegraphics[width=0.45\textwidth]{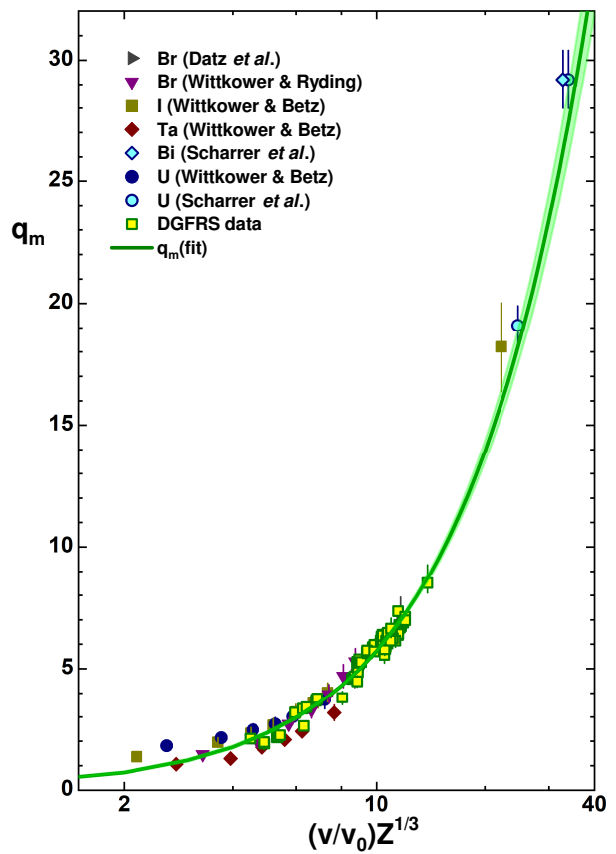}}
\vspace{-3.0mm} \caption{\label{QmMy}Experimental data for mean equilibrated charge of the heavy low energy ions \cite{WitBetzAD73,Schar17} (different symbols) and for the mean charges of ERs obtained with DGFRS \cite{DGFRScha} (open squares) are shown as a function of reduced velocity $(v/v_{0})Z^{1/3}$. The result of the data fit denoted as $q_{m}(\rm fit)$ is also shown (solid line with a shadow area corresponding to the 95\% confidence limit). See details in the text.}
\end{figure}

For the estimates of $q_{m}$ in diluted H$_{2}$, the first inspection was performed of the systematics based on the large body of experimental data obtained for HIs passed through different gas media \cite{SG01}. The accuracy of these $q_{m}^{\rm SG}$ estimates, as claimed in this work, corresponds to $\Delta(q_{m}^{\rm SG}/Z) = \pm2.6\%$. It means that for the heaviest ions with $100 \leqslant Z \leqslant 120$, required values of $q_{m}^{\rm SG}$ are estimated with the accuracy of $\pm$(26--31)\% that is insufficient in many cases. Different scalings of the available $q_{m}$ data for Br and heavier ions \cite{WitBetzAD73,Schar17} together with the Dubna GFRS (DGFRS) data \cite{DGFRScha} were tested and the best one (the least reduced $\chi_{r}^{2}$ value) was obtained for $q_{m}$ being a power function of reduced velocity $(v/v_{0})Z^{1/3}$. In Fig.~\ref{QmMy}, these data are shown together with the result of the $a x^{b}$ function fit, where $x = (v/v_{0})Z^{1/3}$, and $a$ and $b$ are fitted parameters. The magnitudes of $a = 0.2985\pm0.0135$ and $b = 1.283\pm0.020$ were obtained for these parameters (here and below standard errors of parameter values are indicated). This approximation applied to ionized No ERs produced in the $^{48}$Ca+$^{206}$Pb reaction (mean energy $E$ = 39 MeV), which are moving in H$_{2}$, gives us $q_{m}$ = 7.0$\pm$0.1 (for the 95\% confidence limit), i.e., the uncertainty in the $q_{m}$ value is $\pm$1.4\%.

\begin{figure}[!t]
\vspace{0.5mm}
\centerline{\includegraphics[width=0.45\textwidth]{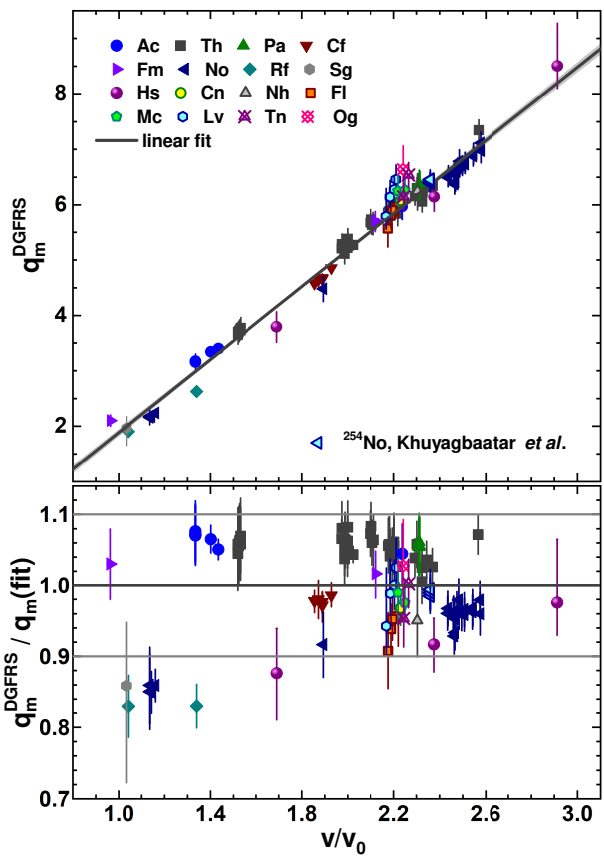}}
\vspace{-3.5mm} \caption{\label{QmDGFRS}Data on mean charges obtained in experiments with DGFRS \cite{DGFRScha} together with new ones and those obtained in \cite{TASCAcha} (symbols) are shown as a function of velocity $v/v_{0}$ (upper panel). The result of a linear fit to the data is also shown (solid line with a shadow area corresponding to the 95\% confidence limit). The same data are compared with the $q_{m}(\rm fit)$ approximation shown in Fig.~\ref{QmMy} (bottom panel).}
\end{figure}

In the experiments with DGFRS, a linear dependence of velocity $v/v_{0}$ for mean charges $q_{m}^{\rm DGFRS}$ was obtained for ERs with $Z \geqslant$ 89 \cite{DGFRScha}. New $q_{m}^{\rm DGFRS}$ data obtained later and the data \cite{TASCAcha} satisfy this dependence, as one can see in Fig.~\ref{QmDGFRS}. All the data correspond to the relationship: $q_{m}^{\rm DGFRS} = (-1.415\pm0.066) + (3.299\pm0.035)v/v_0$, as was obtained with the data fit. The $q_{m}^{\rm DGFRS}$ value for No ERs at $v/v_{0}$ = 2.5 is estimated as 6.83$\pm$0.06 (for the 95\% confidence level). A difference between this estimate and the previous one for the mean charge value leads to the shift in the position of maxima $X$ for ER distributions in the focal plane of DGFRS. This shift corresponds to the relationship: $q_{m}({\rm fit}) / q_{m}^{\rm DGFRS} = 1 + X/(100 D)$. Here $D$ is the dispersion of DGFRS, which is the position shift of the maximum for the ER horizontal focal-plane distribution per one percent change in rigidity $(B\rho)$. In the DGFRS experiments, $D$ was determined to be 7.5 mm \cite{DGFRScha}. Thus, the difference in the $q_{m}$ estimates for No leads the respective shift of $X = 18.7$ mm.

Note that the reduced $\chi^{2}_{r}$ value obtained with the linear fit to the $q_{m}^{\rm DGFRS}$ data against $v/v_{0}$ is much less than the one obtained with a similar fit using $(v/v_{0})Z^{1/3}$ as an argument. The latter was used to systemize the mean-charge data for HIs, fission fragments, and ERs in He \cite{Gregor13}. In the bottom panel of Fig.~\ref{QmDGFRS}, the DGFRS data are compared with the $q_{m}(\rm fit)$ approximation shown in Fig.~\ref{QmMy}. As one can see, 92\% of the $q_{m}^{\rm DGFRS}$ data points are spread within a corridor, corresponding to the $\pm$10\% deviation of the approximation. Such spread is very similar to the one observed in the analysis of $q_{m}$ data obtained in He gas (see Fig.~2 in \cite{Gregor13}). At the same time, a better data fit was obtained taking into account the `electronic shell correction' $\delta q_{m}$ \cite{Gregor13} (see Figs.~4 and 8 herein). The amplitude of this correction for $q_{m}^{\rm DGFRS}$ corresponds to the value of $\delta q_{m}^{\rm DGFRS}$ = $\pm$0.3, which is similar to the data point errors (see Fig.~\ref{QmDGFRS} and Fig.~8 in \cite{Gregor13}). In further considerations, the `shell correction' for $q_{m}^{\rm DGFRS}$ was neglected.

One would think that charge-changing cross-section data available for HIs (not for ERs) allow us to obtain mean equilibrated charges similar to those shown in Fig.~\ref{QmMy}, despite the difference in experimental conditions used for the retrieval of these data. Cross-section data can also be used for the description of the charge equilibration for ERs. A joint presentation of mean equilibrated charges obtained for HIs and ERs shown in Fig.~\ref{QmMy} and their comparisons in Fig.~\ref{QmDGFRS} confirm this statement. A similar data presentation was also used for mean charges in He gas (see Fig.~2 in \cite{Ghio88,Gregor13}). Based on this assertion, different approximations to experimental charge-changing cross-sections were used in a number of works considering the transmission of ERs through a rarefied gas of gas-filled separators \cite{Ninov95,Paul89,Gregor13,Saren11}.

\section{\label{CSparam}Charge-changing cross sections and their approximations}

As was mentioned in Section \ref{MeanChar}, single-electron capture and loss cross-sections, as exponential functions of the difference between HI charge state $q$ and fixed value of $q_{m}$, were usually used for the consideration of a charge-changing process inside a gas. Assuming that an equilibrated charge distribution has a Gaussian shape with mean value $q_{m}$ and standard deviation $d$, appropriate exponential functions can be written using these values \cite{Ninov95,Betz73,Paul89,Gregor13}. Effective parameters of the exponential functions for capture and loss cross-sections $\sigma_{q,q-1}$ and $\sigma_{q,q+1}$, respectively, are interconnected \cite{Betz73,Paul89} if $q_{m}$ and $d$ are used. The lasts could be taken from the empirical systematics \cite{Betz73,WitBetz73,DmiNik64}. Parameters of the $\sigma_{q,q-1}$ function were derived with a fit to the cross-sections given by the analytical formulae \cite{Knud81,Schla83} and applied in respective simulations for $^{252}$No and $^{259}$Rf ERs passing through He \cite{Gregor13} and for $^{58}$Ni ions passing through N$_{2}$ \cite{Paul89}.

\subsection{\label{CScapcomp}Testing approximations for electron capture cross sections}

\begin{figure}[!t]
\vspace{0.0mm}
\centerline{\includegraphics[width=0.45\textwidth]{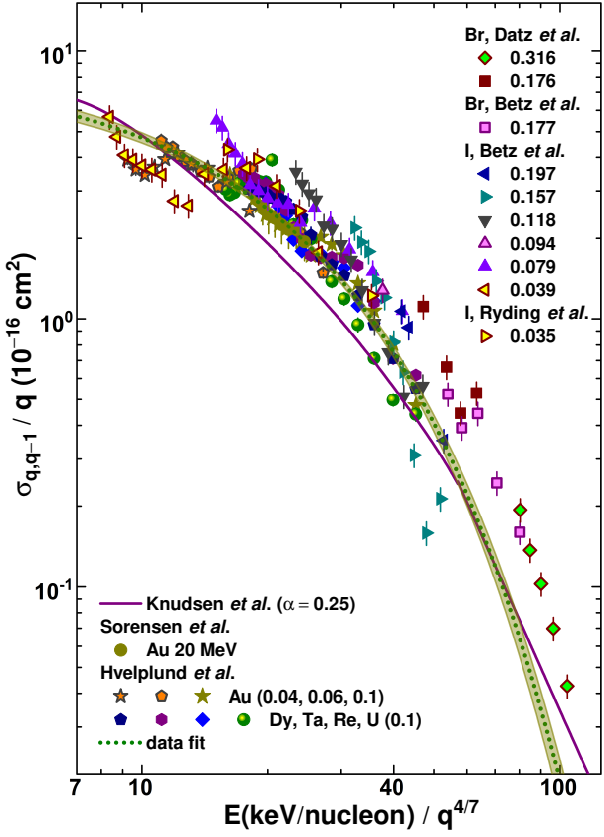}}
\vspace{-3.5mm} \caption{\label{Knudcomp}Reduced capture cross-sections $\sigma_{q,q-1}/q$ as obtained in the experiments \cite{Betz73,Soren84,Hvelp92} for low energy heavy ions (different symbols) with charge state $q$ and energy $E$ (in keV/nucleon), which have passed through H$_{2}$, in comparison with the semi-empirical dependence \cite{Knud81} for $\alpha = 0.25$, which is the adjustable parameter with a value between 0 and 1 (solid line). The result of the exponential function fit to the data is shown by a dotted curve with a shadow area corresponding to the 95\% confidence limit. See the text for details.}
\end{figure}

\begin{figure}[!h]
\vspace{-0.5mm}
\centerline{\includegraphics[width=0.45\textwidth]{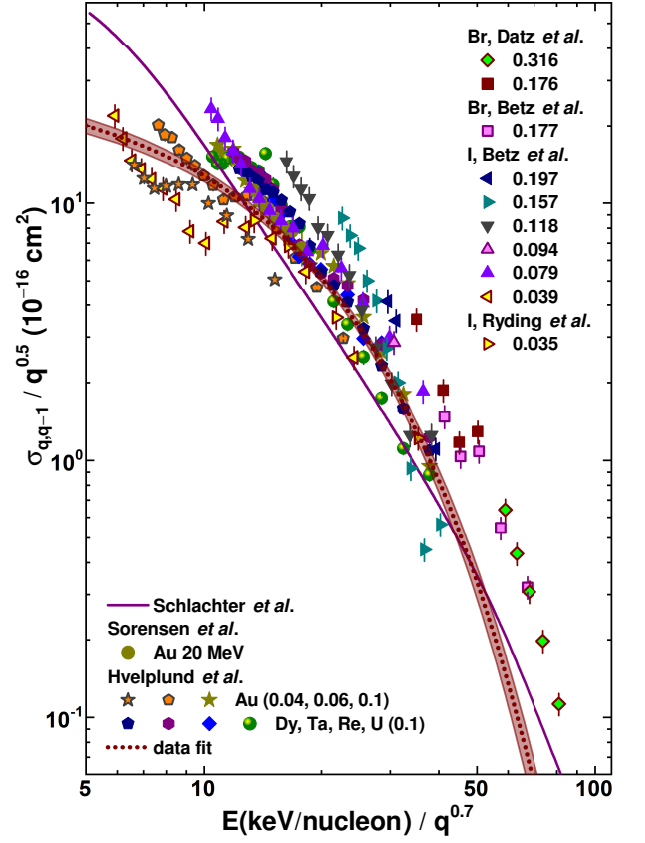}}
\vspace{-3.5mm} \caption{\label{Schlacomp}The same as in Fig.~\ref{Knudcomp}, but for reduced capture cross-sections $\sigma_{q,q-1}/q^{0.5}$ as a function of reduced energy $E/q^{0.7}$ in comparison with the empirical formula \cite{Schla83}.}
\end{figure}

Compatibility between $\sigma_{q,q-1}$ obtained for low energy heavy ions passed through H$_{2}$ \cite{Betz73,Soren84,Hvelp92} and the same values resulted in the approximation \cite{Knud81} was checked to get an idea of their applicability for simulations of the charge-changing process in the lightest gas. One can remind that the approximation \cite{Knud81} is based on the notion of a classical cross-section introduced by Bohr and Lindhard \cite{BohrLind54}. The same test for applicability of the empirical formula \cite{Schla83} was also performed. The results of both of these inspections are shown in Figs.~\ref{Knudcomp} and \ref{Schlacomp}.

\begin{figure}[!t]
\vspace{0.5mm}
\centerline{\includegraphics[width=0.45\textwidth]{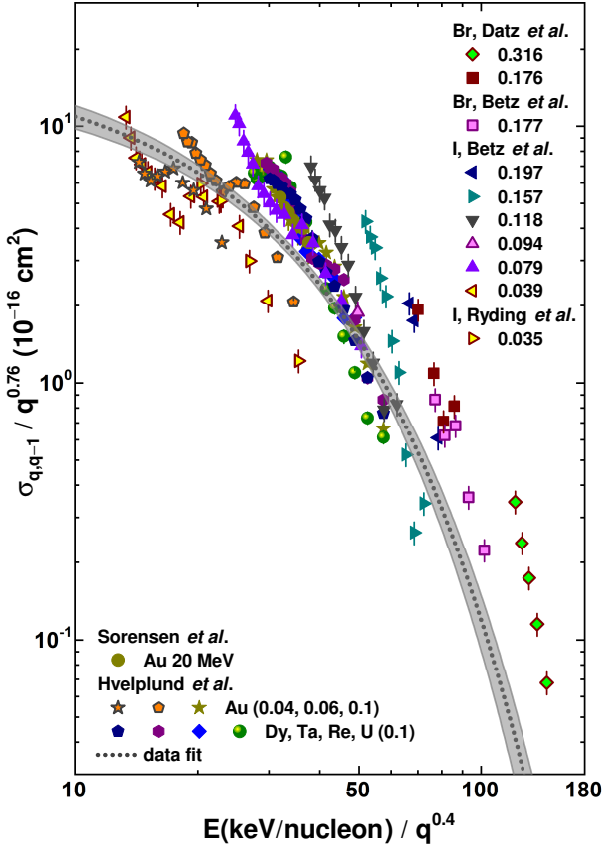}}
\vspace{-3.5mm} \caption{\label{Corncomp}The same as in Figs.~\ref{Knudcomp} and \ref{Schlacomp}, but for reduced capture cross-sections $\sigma_{q,q-1}/q^{0.76}$ as a function of reduced energy $E/q^{0.4}$ \cite{Cornel06}.}
\end{figure}

Besides the approximations \cite{Knud81,Schla83}, a single universal scaling rule was later proposed with the use of reduced capture cross-section $\sigma_{q,q-1}/q^{0.76}$ as a function of reduced energy $E/q^{0.4}$ \cite{Cornel06}. This scaling is `extremely well defined for $E/q^{0.4} > 5$ keV/nucleon and includes all projectile charge states', as was mentioned in this work. According to Fig.~1 in \cite{Cornel06}, perfect data scaling is demonstrated, but for HIs not heavier than Fe. Thus, the applicability of this scaling rule has to be also tested for possible usage in simulations. In Fig.~\ref{Corncomp}, the result of the application of the scaling rule \cite{Cornel06} to the same cross-section data, as used in Figs.~\ref{Knudcomp} and \ref{Schlacomp}, is shown. As one can see, this scaling gives appreciably worse results than those proposed in \cite{Knud81,Schla83}.

Note that in the subsequent application of any approximation, it should be taken into account that initial charge state distributions for heavy ERs are assumed to be very broad, as mentioned in Section~\ref{intro}. For example, for $^{252}$No ERs produced in the $^{206}$Pb($^{48}$Ca,2$n$) reaction, which are knocked out from a relatively thin target with the mean energy of 39 MeV, the initial charge states cover the range of +10 $\leqslant q \leqslant$ +80, as will be further considered in Section~\ref{MCsim}. Under these conditions, the appropriate reduced energies are in the regions of 42 $\geqslant E/q^{4/7} \geqslant$ 13, 31 $\geqslant E/q^{0.7} \geqslant$ 7.2, and 62 $\geqslant E/q^{0.4} \geqslant$ 27 for the scalings of \cite{Knud81}, \cite{Schla83}, and \cite{Cornel06}, respectively. Single-electron capture cross-sections corresponding to these regions of reduced energies, which were obtained with the exponential function fits to the data, vary in the ranges of (6.9$\pm$0.4) $\leqslant \sigma_{q,q-1} \leqslant$ (323$\pm$12), (6.0$\pm$0.4) $\leqslant \sigma_{q,q-1} \leqslant$ (147$\pm$9), and (4.7$\pm$0.5) $\leqslant \sigma_{q,q-1} \leqslant$ (131$\pm$8) in the units of 10$^{-16}$ cm$^{2}$ with the indicated errors corresponding to the 95\% confidence level, as shown in Figs.~\ref{Knudcomp}--\ref{Corncomp}. The transition from scaling \cite{Knud81} to \cite{Cornel06}, corresponds to the degradation of the quality of fits (increase in $\chi^{2}_{r}$ values) accompanied by increase in relative uncertainties of the cross-sections. Hereinafter, the exponential approximation for single-electron capture cross-sections in the form $a_{c}\exp(b_{c} E_{r})$, where $E_{r} = E/q^{4/7}$ \cite{Knud81}, with the use of the fitted parameters $a_{c} = 8.77 \pm 0.29$, and $b_{c} = -0.0614 \pm 0.0013$, will be used in simulations of charge-changing process for heavy ERs.

\subsection{\label{CSFrancomp}Empirical approximations of B. Franzke \cite{Franz81}}

In 1981, B.~Franzke proposed two empirical formulae for single-electron capture and loss cross-sections \cite{Franz81}. They are a power function of $q/q_{m}$ with the positive and negative index of power $p$ for $\sigma_{q,q-1}$ and $\sigma_{q,q+1}$, respectively. The value of $p$ depends on the ratio $q/q_{m}$. Good agreement between experimental data and calculated values was shown for the loss cross-sections of 1.4 MeV/nucleon U ions colliding with N$_{2}$. At the same time, no comparison for capture cross-sections was presented, and hence testing these formulae was of interest for their possible application.

\begin{figure}[!t]
\vspace{0.5mm}
\centerline{\includegraphics[width=0.45\textwidth]{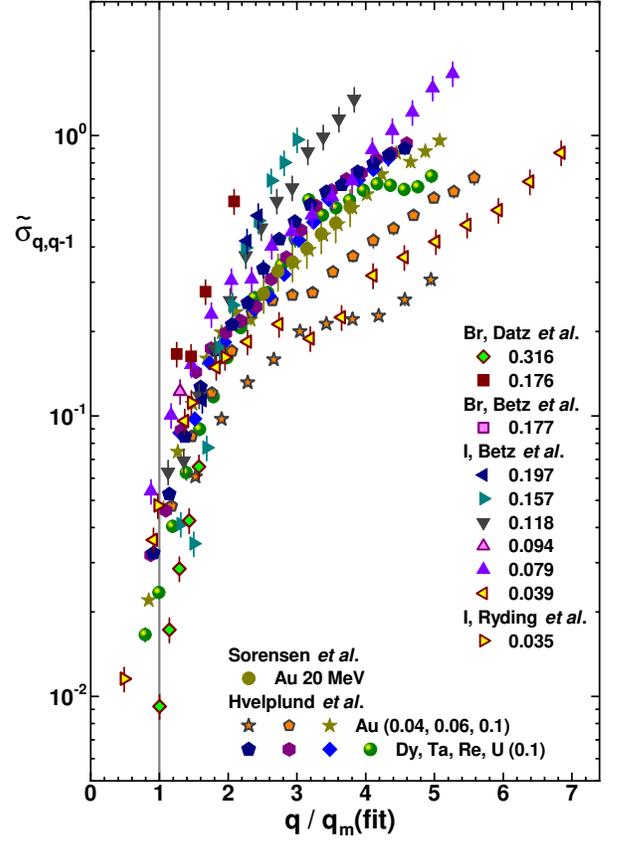}}
\vspace{-3.5mm} \caption{\label{Francapcomp}The same as in Figs.~\ref{Knudcomp}, \ref{Schlacomp}, and \ref{Corncomp} but for reduced capture cross-sections $\tilde{\sigma}_{q,q-1}$ according to Eq.~(\ref{sicapred}), which is a function of $q/q_{m}(\rm fit)$, where $q_{m}(\rm fit)$ is a power function fitted to the $q_{m}$ data shown in Fig.~\ref{QmMy}. See the text for details.}
\end{figure}

According to Eq.~(1) presented in \cite{Franz81}, the electron capture cross-section, omitting some factors, is approximated by a function of $(q/q_{m})^{p}$, where $p$ depends on $q$ ($p$ = 4 for $q \leqslant q_{m}$ and $p$ = 2 for $q \geqslant q_{m}$). For low energies (in a non-relativistic approximation), reduced cross-section $\tilde{\sigma}_{q,q-1}$ transformed from Eq.~(1) \cite{Franz81} becomes a function of $(q/q_{m})^{p}$ and can be written as
\begin{equation}
\label{sicapred}
 \tilde{\sigma}_{q,q-1} = 230.5 E^{2} \sigma_{q,q-1} / q_{m}^{2} / Z^{1/2},
\end{equation}
where $\sigma_{q,q-1}$ is the experimental cross-section expressed in the units of 10$^{-16}$ cm$^{2}$, and $E$ is the HI energy in MeV/nucleon. An attempt was made to get the data scaling with the reduced cross-section values determined by Eq.~(\ref{sicapred}) as a function of $q/q_{m}$ and the same $\sigma_{q,q-1}$ data as was presented in Figs.~\ref{Knudcomp}, \ref{Schlacomp}, and \ref{Corncomp}. Such data representation is shown in Fig.~\ref{Francapcomp} for $q/q_{m}(\rm fit)$ as an argument, where $q_{m}(\rm fit)$ is a power function shown in Fig.~\ref{QmMy}. As one can see, $\tilde{\sigma}_{q,q-1}$ values thus obtained radically diverge from each other at $q/q_{m}(\rm fit)>1.5$ for HIs with different masses, charge states and energies. Therefore, it can be stated that Eq.~(\ref{sicapred}) should be modified to ensure better data scaling. A similar divergence was revealed for the $q_{m}$ values corresponding to the linear approximation of the $q_{m}^{\rm DGFRS}$ data shown in Fig.~\ref{QmDGFRS}.

For single-electron loss cross-sections, an empirical formula proposed by the author (Eq.~(2) in \cite{Franz81}) was also tested. It seems more convenient for practical routine simulations than more elaborated approaches \cite{Sant04,Shev12}. For low energies (in a non-relativistic approximation), reduced single-electron loss cross-section $\tilde{\sigma}_{q,q+1}$, according to Eq.~(2) in \cite{Franz81}, can be written as
\begin{equation}
\label{silossred}
 \tilde{\sigma}_{q,q+1} = 1.3239 E^{1/2} q_{m}^{2} \sigma_{q,q+1} / 10^{Y},
\end{equation}
where $\sigma_{q,q+1}$ is expressed in the units of 10$^{-16}$ cm$^{2}$, and $Y = (0.71 \log Z)^{3/2}$ is the parameter deduced from calculated binding energies for the outermost electrons \cite{Franz81}. As in the case of capture cross-sections, the reduced cross-section is a function of $(q/q_{m})^{p}$, in which index $p$ depends on $q$ ($p = -$2.3 for $q \leqslant q_{m}$ and $p = -$4 for $q \geqslant q_{m}$ \cite{Franz81}).

\begin{figure}[!t]
\vspace{0.5mm}
\centerline{\includegraphics[width=0.45\textwidth]{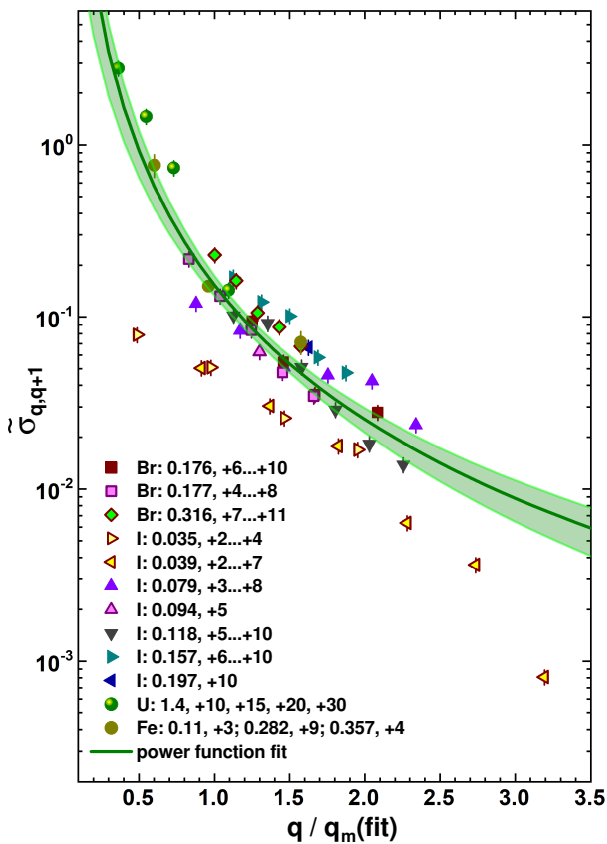}}
\vspace{-3.5mm} \caption{\label{CSlossQmy}Reduced single-electron loss cross-sections $\tilde{\sigma}_{q,q+1}$ obtained with Eq.~(\ref{silossred}) applied to the data for low energy heavy ions passed through H$_{2}$. Heavy ions designated by different symbols have the energies (in MeV/nucleon) and charge states indicated in the figure \cite{Betz73,Franz81,Berk81}. Mean equilibrated charges $q_{m}(\rm (fit)$ were calculated according to the function shown in Fig.~\ref{QmMy}. The result of a power function fit to the data is shown by a solid curve with a shadow area corresponding to the 95\% confidence limit.}
\end{figure}

Unfortunately, it is a lack of available data on single-electron loss cross-sections, corresponding to HIs colliding with H$_{2}$ \cite{Betz73}, which have appropriate masses and energies. For the analysis, the data for U ions at rather high energy of 1.4 MeV/nucleon \cite{Franz81} and for Fe ions at $E \leqslant$ 0.357 MeV/nucleon \cite{Berk81} were added to the data compiled in \cite{Betz73}. The experimental cross-sections \cite{Betz73,Franz81,Berk81} were converted into reduced ones using Eq.~(\ref{silossred}), and such scaling was tested with the empirical $q_{m}$ values resulted in the approximations shown in Figs.~\ref{QmMy} and \ref{QmDGFRS}. The $\tilde{\sigma}_{q,q+1}$ values were plotted as a function of $q/q_{m}$, and fitted with a power function $a_{pl}(q/q_{m})^{b_{pl}}$, as shown in Fig.~\ref{CSlossQmy}. The best fit to the data (the least $\chi^{2}_{r}$ values) was obtained with the $q_{m}(\rm fit)$ approximation shown in Fig.~\ref{QmMy}. The iodine cross-section data at $E <$ 0.04 MeV/nucleon (see \cite{Betz73} and Fig.~\ref{CSlossQmy}) were ignored in the fitting procedure since they significantly deviated from the general dependency $\tilde{\sigma}_{q,q+1} = f(q/q_{m})$ observed for the tested $q_{m}$ values. Parameter values obtained with the power function fit to the $\tilde{\sigma}_{q,q+1}$ data using the $q_{m}(\rm fit)$ values are as follows: $a_{pl} = 0.153\pm0.010$ and $b_{pl} = -2.59\pm0.15$. For $^{252}$No at the highest initial charge state  $q \simeq +80$, the $\sigma_{q,q+1}$ value thus obtained corresponds to 3$\cdot$10$^{-19}$ cm$^{2}$, which is less than the respective $\sigma_{q,q-1}$ value by several orders of magnitude. The loss cross-sections become comparable with the capture ones at the charge states from +3 to +10, i.e., in the vicinity of the equilibrated charge. At these charge states, the electron loss cross-sections vary in the region of (15$\pm$4) $\geqslant \sigma_{q,q+1} \geqslant$ (0.68$\pm$0.07) in the units of 10$^{-16}$ cm$^{2}$ with the errors corresponding to the 95\% confidence level, as shown in Fig.~\ref{CSlossQmy}.

\subsection{\label{newappCSloscap}New empirical approximations for charge-changing cross-sections}

Scaling of experimental data according to the ratio of $q/q_{m}$ could be the right way for a general description of charge-changing cross-sections, bearing in mind its quite acceptable application to the estimates of electron loss cross-sections (see Section~\ref{CSFrancomp}). In this context, it was revealed that a reasonable scaling for electron capture cross-sections could be obtained with $q/q_{m}$ earlier used as an argument. A power function and the exponential one in the form of $a_{ec}\exp\{b_{ec}/[(q/q_{m}) + c_{ec}]\}$ were used to fit the data. Fits with the exponential function using the $q/q_{m}^{\rm DGFRS}$ ratio showed better results than those obtained with the power one for any choice of $q/q_{m}$ (the least $\chi^{2}_{r}$ values were obtained with the exponent).

\begin{figure}[h!]
\vspace{0.5mm}
\centerline{\includegraphics[width=0.45\textwidth]{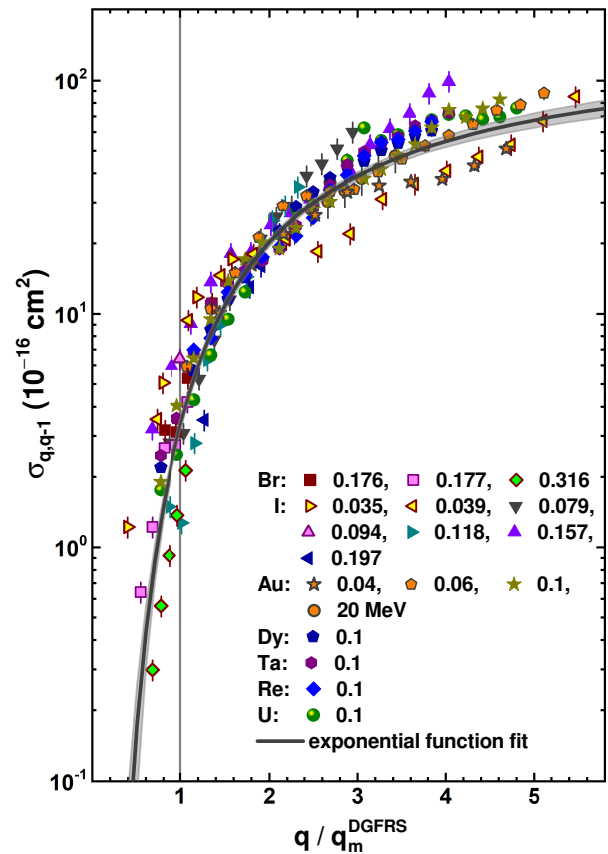}}
\vspace{-3.5mm} \caption{\label{CScapQQm}Electron capture cross-sections $\sigma_{q,q-1 }$ for low energy heavy ions passed through H$_{2}$ \cite{Betz73,Soren84,Hvelp92} are shown as a function of $q/q_{m}^{\rm DGFRS}$ (symbols with respective energies indicated in MeV/nucleon) for $q_{m}^{\rm DGFRS}$ calculated according to the results of data fit shown in Fig.~\ref{QmDGFRS}. The result of an exponential function fit to the data is shown by a solid curve with a shadow area corresponding to the 95\% confidence limit. See the text for details.}
\end{figure}

\begin{table*}[ht!]
\caption{\label{sumpar} Parameter values $a$, $b$, and $c$ of the empirical formulae obtained for determination of the equilibrated mean charge $q_{m}$, electron capture, and loss cross-sections, $\sigma_{q,q-1}$ and $\sigma_{q,q+1}$, respectively, with the use of specified argument $x$. With these parameter values, $E$ in MeV/nucleon and $Y = [0.71\log(Z)]^{3/2}$, $\sigma_{q,q-1}$ and $\sigma_{q,q+1}$ are obtained in the units of 10$^{-16}$ cm$^{2}$.}
\begin{ruledtabular}
\renewcommand{\arraystretch}{1.25}
\begin{tabular}{cccccc} \\ \hline
Estimated & Empirical & Argument         & \multicolumn{3}{c}{Parameter values} \\ \cline{4-6}
value     & formula   &        $x$           &       $a$      &   $b$  &  $c$                 \\ \hline
$q_{m}(\rm fit)$   &   $ax^{b}$    &$(v/v_{0})Z^{1/3}$ &  0.2985 & 1.283  &     \\
$q_{m}^{\rm DGFRS}$& $a + b x$     &    $v/v_{0}$      &$-$1.415 & 3.299  &     \\
$\sigma_{q,q-1}$   & $aq\exp(-bx)$ & $1000E/q^{4/7}$   &   8.77  & 0.0614 &     \\
                   & $a\exp[-b/(x+c)]$& $q/q_{m}^{\rm DGFRS}$  & 160    & 4.44 & 0.15 \\
$\sigma_{q,q+1}$   &$0.755ax^{b}10^{Y}\!/\!E^{1/2}\!/q_{m}^{2}$&$q/q_{m}(\rm fit)$    &0.153& $-$2.59 &     \\
$\sigma_{q,q+1}/\sigma_{q,q-1}$  & $ax^{b}$&$q/q_{m}(\rm fit)$ & 0.345  & $-$4.32 & \\
\end{tabular}
\end{ruledtabular}
\end{table*}

In Fig.~\ref{CScapQQm}, the electron capture cross-section data \cite{Betz73,Soren84,Hvelp92} are shown as a function of $q/q_{m}^{\rm DGFRS}$ corresponding to the best data scaling. Parameter values of the exponential function relating to this fit are as follows: $a_{ec} = 160\pm14$, $b_{ec} = -4.44\pm0.34$ and $c_{ec} = 0.150\pm0.065$. Single-electron capture cross-sections for ionized No ERs with +10 $\leqslant q \leqslant$ +80, which are obtained with this approximation, vary in the region of (10.3$\pm$0.5) $\leqslant \sigma_{q,q-1} \leqslant$ (116$\pm$12) (in the units of 10$^{-16}$ cm$^{2}$ and with errors corresponding to the 95\% confidence level, as shown in Fig.~\ref{CScapQQm}). As one can see, these estimates somewhat differ from the previous ones obtained with approximations using scalings according to \cite{Knud81}, \cite{Schla83}, and \cite{Cornel06} (see Section~\ref{CScapcomp}).

\begin{figure}[b!]
\vspace{-0.5mm}
\centerline{\includegraphics[width=0.45\textwidth]{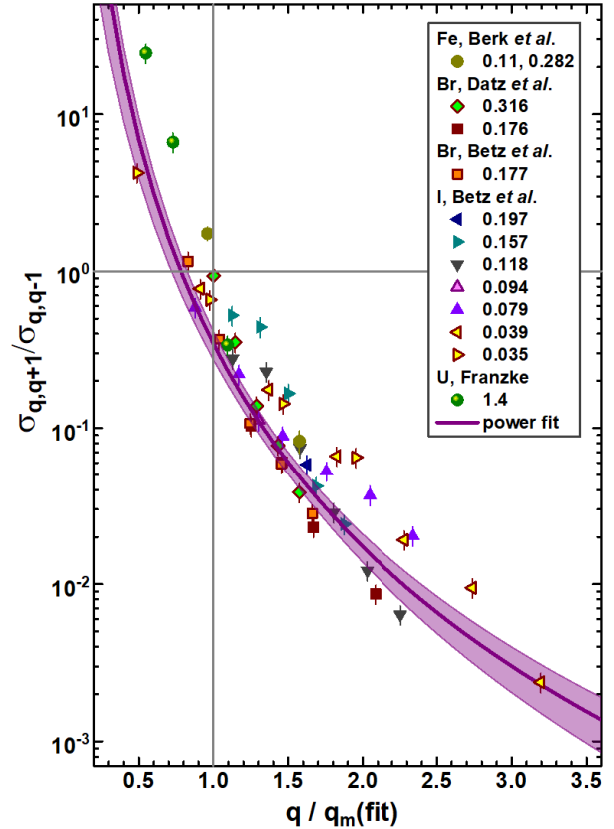}}
\vspace{-3.0mm} \caption{\label{CSlocapQQm}The ratios of $\sigma_{q,q+1}/\sigma_{q,q-1}$ for low energy heavy ions passed through H$_{2}$ \cite{Betz73,Franz81,Berk81} are shown as a function of  $q/q_{m}(\rm fit)$ calculated according to the results of data fit shown in Fig.~\ref{QmMy} (symbols with respective energy values indicated in MeV/nucleon). The result of a power function fit to the data is shown by a solid curve with a shadow area corresponding to the 95\% confidence limit.}
\end{figure}

Simple scaling was also found for the single-electron loss cross-sections expressed with ratio $\sigma_{q,q+1}/\sigma_{q,q-1}$ as a function of $q/q_{m}$ in the application of this approach to the data \cite{Betz73,Franz81,Berk81}. As in the previous cases, two variants with the $q_{m}(\rm fit)$ and $q_{m}^{\rm DGFRS}$ mean charges were tested. The best scaling was obtained with the $q_{m}(\rm fit)$ charges, according to the results of fits using the exponential function $a_{el}\exp(b_{el}x)$ and power one $a_{pl}x^{b_{pl}}$ for $x = q/q_{m}$. The result of the power function fit, corresponding to the least $\chi^{2}_{r}$ value, is shown in Fig.~\ref{CSlocapQQm}. Parameter values for this function are as follows: $a_{pl} = 0.345\pm0.036$, $b_{pl} = -4.32\pm0.19$. As in the previous case, dealing with the $^{252}$No reduced cross-sections (see Fig.~\ref{CSlossQmy}), single-electron loss cross-sections at the highest initial charge states are much less than the respective capture ones. The $\sigma_{q,q+1}$ values become comparable with the $\sigma_{q,q-1}$ ones in the vicinity of the equilibrated charge, i.e., at the charge states from +3 to +10. At these charge states, according to the approximation, the electron loss cross-sections vary in the region of (2.5$\pm$1.2) $\geqslant \sigma_{q,q+1} \geqslant$ (0.55$\pm$0.09) (in the units of 10$^{-16}$ cm$^{2}$ and with errors corresponding to the 95\% confidence level shown in Fig.~\ref{CSlocapQQm}). These estimates correspond to the capture cross-section approximation shown in Fig.~\ref{Knudcomp}.

The empirical scalings considered in this Section and Section~\ref{MeanChar} can be used for the estimates of the equilibrated mean charge, single-electron capture and loss cross-sections in rarefied H$_{2}$ for low energy HIs and for ERs produced in fusion-evaporation reactions. The results of these approximations are summarized in Table~\ref{sumpar}.

\begin{figure}[h!]
\vspace{0.5mm}
\centerline{\includegraphics[width=0.45\textwidth]{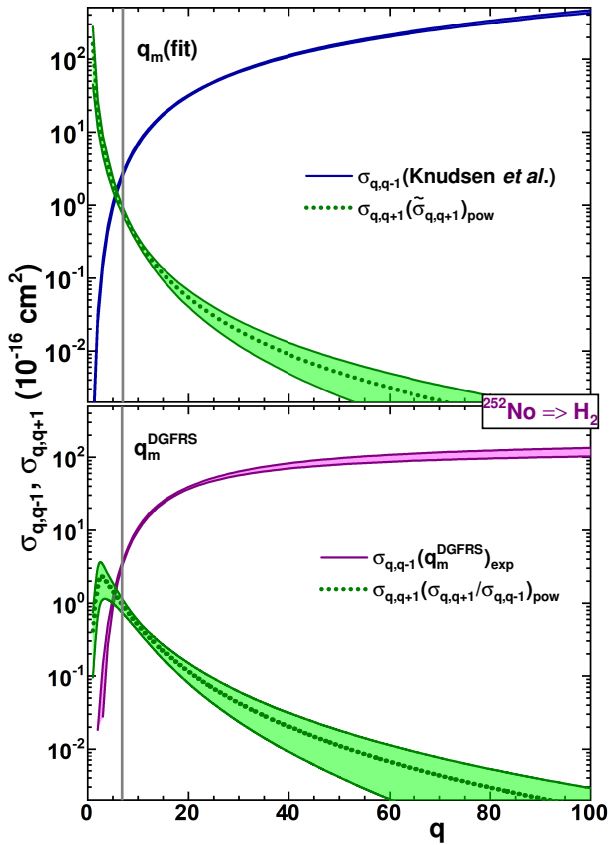}}
\vspace{-3.0mm} \caption{\label{252NoCS}Single-electron capture $\sigma_{q,q-1}$ and loss $\sigma_{q,q+1}$ cross-sections for the 39 MeV $^{252}$No ions moving in H$_{2}$ are shown as a function of ion charge state $q$ (lines with a shadow area corresponding to the 95\% confidence limit). The cross section scalings shown in Figs.~\ref{Knudcomp} and \ref{CSlossQmy}, and in Figs.~\ref{CScapQQm} and \ref{CSlocapQQm} are used for the calculations presented in the upper and bottom panel, respectively. See details for the scaling functions in Table~\ref{sumpar}.}
\end{figure}

The approximations presented in Table~\ref{sumpar}, being applied to $^{252}$No ions at their charge states on the edge of the assumed region, showed a significant spread in the cross-section values, as mentioned above. In Fig.~\ref{252NoCS}, single-electron capture and loss cross-sections are shown as functions of charge states $q$ assumed for the 39 MeV ionized $^{252}$No ERs. This mean energy is inherent in the ERs produced in the $^{206}$Pb($^{48}$Ca,2$n$) reaction induced by the 217 MeV $^{48}$Ca beam. The calculations performed with the parameter values corresponding to the best data fits obtained with exponential and power functions presented in Table~\ref{sumpar}. As one can see in Fig.~\ref{252NoCS}, empirical mean equilibrated charges $q_{m}(\rm fit)$ and $q_{m}^{\rm DGFRS}$ used in these approximations differ from those corresponding to $\sigma_{q,q-1} = \sigma_{q,q+1}$. Some explanations of the difference that is also seen in Fig.~\ref{CSlocapQQm} one can find in \cite{Betz73}.

\section{\label{MCsim}Monte Carlo simulations of charge-changing process for ERs}

\begin{figure}[h!]
\vspace{0.5mm}
\centerline{\includegraphics[width=0.475\textwidth]{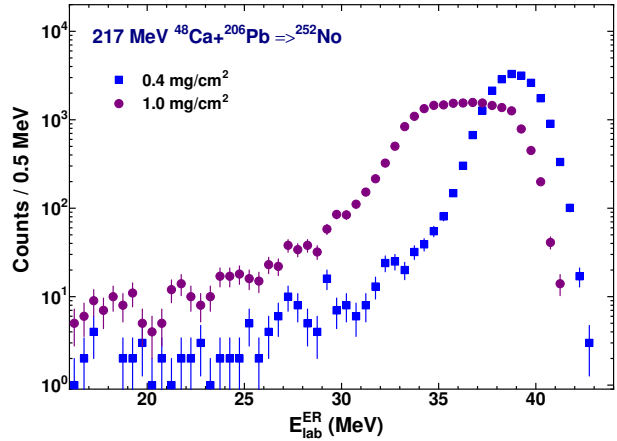}}
\vspace{-3.0mm}
\caption{\label{EdisNo}The energy distributions for $^{252}$No produced in the $^{206}$Pb($^{48}$Ca,$2n$) reaction at the input energy of 217 MeV for the target thicknesses of 0.4 and 1.0 mg/cm$^{2}$ (squares and circles, respectively), as obtained in the MC simulations \cite{SagaNIM13}.}
\end{figure}

\begin{figure}[h!]
\vspace{0.5mm}
\centerline{\includegraphics[width=0.475\textwidth]{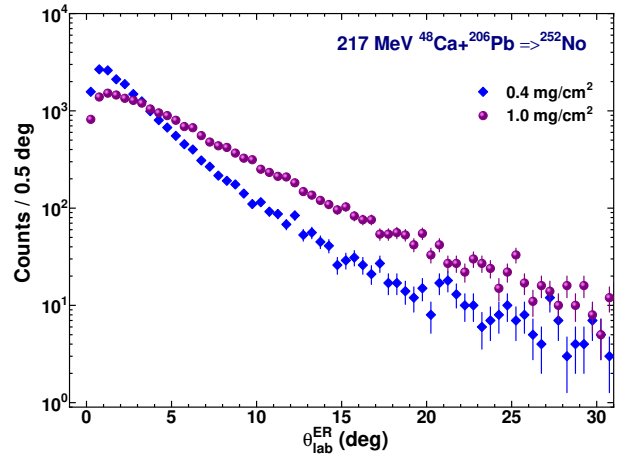}}
\vspace{-2.0mm}
\caption{\label{AndiNo}The angular distributions for $^{252}$No produced in the $^{206}$Pb($^{48}$Ca,$2n$) reaction at the input energy of 217 MeV for the target thicknesses of 0.4 and 1.0 mg/cm$^{2}$ (diamonds and spheres, respectively), as obtained in the MC simulations \cite{SagaNIM13}.}
\end{figure}

ERs knocked out from relatively thin targets by a heavy ion beam have well-determined energy distributions and forward peaked angular distributions, as the results of their straight-forward production in the complete fusion-evaporation reactions. These distributions can be obtained with Monte Carlo (MC) simulations taking into account the evaporation of light particles from compound nuclei formed in the reaction and processes of stopping and multiple scattering of ER atoms inside a target \cite{SagaNIM13}. In Figs.~\ref{EdisNo} and \ref{AndiNo}, the energy and angular distributions for $^{252}$No produced in the $^{206}$Pb($^{48}$Ca,2n) reaction are shown, which were obtained as described earlier \cite{SagaNIM13}. The energy distributions will be used for the MC simulations of charge distributions for No ions escaped the targets, whereas the angular distributions are given for the reference.

According to the previous considerations \cite{Ninov95,Paul89,Gregor13}, a distance of HI between successive charge-changing collisions is determined as
\begin{equation}
\label{pathlength}
  l = -\lambda \ln P,
\end{equation}
where $\lambda$ is the mean free path between two charge-changing collisions and $P$ is the probability for HI to survive flight path $l$ without any collisions (a random number between 0 and 1). The mean free path $\lambda$ is related to the total charge-changing cross-section $\sigma_{tot} = \sigma_{q,q-1} + \sigma_{q,q+1}$ according to the relation:
\begin{equation}
\label{lambda}
  \lambda = 1 / (n\sigma_{tot}),
\end{equation}
where $n$ is the molecular density of a gas in cm$^{-3}$. For the $\sigma_{q,q-1}$ and $\sigma_{q,q+1}$ cross-sections, one can use the suitable approximations obtained in Section~\ref{CSparam}. The multiple electron capture and loss processes are neglected. The absolute cross-section values for $^{252}$No ions moving in rarefied H$_{2}$ are shown as a function of their charge states in Fig.~\ref{252NoCS}. Simple estimates with these cross-section values and Eq.~(\ref{lambda}) show that for the H$_{2}$ molecular density of 3.3$\cdot$10$^{16}$ cm$^{-3}$ at the pressure of 1 Torr, $\lambda$ increases from 2.5$\cdot$10$^{-4}$ cm ($q$ = +80) to 7.5$\cdot$10$^{-2}$ cm ($q$ = +5), as $\sigma_{tot}$  decreases from 3$\cdot$10$^{-14}$ to 3$\cdot$10$^{-16}$ cm$^{2}$. Cross-sections shown in Fig.~\ref{252NoCS} will be further used in simulations of the charge state distributions for No ERs.

\begin{figure}[t!]
\vspace{0.5mm}
\centerline{\includegraphics[width=0.475\textwidth]{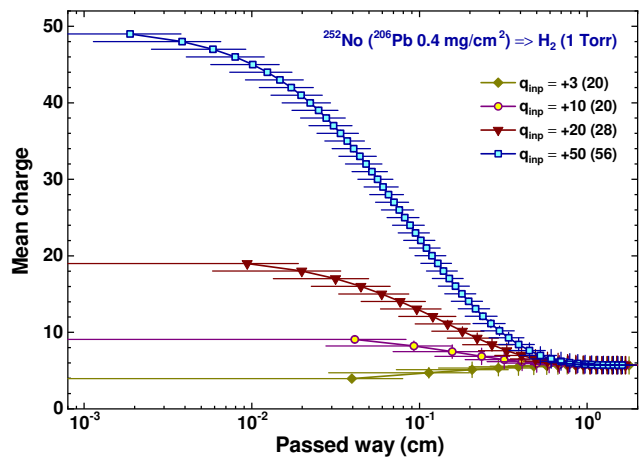}}
\vspace{-2.0mm}
\caption{\label{Nofixchar}Mean charges of $^{252}$No ions as obtained in MC simulations at each step of charge-changing collisions (starting from the first one) are shown as a function of the mean passed way (symbols connected by solid lines) for different input charge state $q_{\rm inp}$. Cross-sections presented in Fig.~\ref{252NoCS} (upper panel) were used in these simulations. Vertical and horizontal bars in the points correspond to standard deviations obtained for the charge and passed way, respectively. The number of collisions leading to the mean equilibrated charge is indicated in parens. See details in the text.}
\end{figure}

In Fig.~\ref{Nofixchar}, the evolution of the input charge states for $^{252}$No ions knocked out from the 0.4 mg/cm$^{2}$ $^{206}$Pb target into the H$_{2}$ gas under the pressure of 1 Torr is shown. Mean charges $q_{m}$ and corresponding mean passed ways $L_{m}$ were obtained with the statistical analysis of charge-state and passed way distributions resulted in MC simulations. The respective charge-changing cross-sections shown in Fig.~\ref{252NoCS} (upper panel) and the energy distribution (Fig.~\ref{EdisNo}) were used in these simulations. Mean charges are shown as a function of $L_{m}$ increasing with the number of the charge-changing collisions (starting from the first one). As one can see, at $L_{m} \gtrsim 1.5$ cm, $q_{m}$ becomes equal to 5.75 irrespective of the input charge state. This value is achieved with the number of the collisions becoming as higher as the input charge state increases. Note that the mean charge value, corresponding to $\sigma_{q,q+1} = \sigma_{q,q-1}$, is equal to 5.65 (see Fig.~\ref{252NoCS}), which slightly differs from the $q_{m}$ value obtained in simulations. Such a difference is the result of different slops in $\sigma_{q,q+1}(q)$ and $\sigma_{q,q-1}(q)$ cross sections \cite{Betz73}.

The value of the passed way thus obtained, which corresponds to the establishment of a charge equilibration, can be compared with a similar one observed in an experiment. For example, for the 13.9 MeV Br ions with the charge states +6, +7, +8, and +10, a charge equilibration is maintained at the H$_{2}$ target thickness of $W_{t} \gtrsim$ 4.5$\cdot$$10^{16}$ mol/cm$^{2}$ (see Fig.~3 in \cite{Datz70}). This value is close to the one obtained in the present simulations ($W_{t} \gtrsim$ 4.9$\cdot$$10^{16}$ mol/cm$^{2}$), corresponding to $L_{m} \gtrsim 1.5$ cm at the H$_{2}$ molecular density of 3.3$\cdot$$10^{16}$ cm$^{-3}$.

In further consideration, it was assumed that the reliable initial charge state distribution for ionized $^{252}$No ERs escaping the $^{208}$Pb target is the composition of the `normal solid' component and of the non-equilibrated one. The `normal solid' component corresponds to the equilibrated charge-state distribution for HIs passed through a solid \cite{SIM82,SY94,SG01}, whereas the non-equilibrated component has charges which are much higher than `normal' ones \cite{Steig63,Scob83,Brink94,Saga08,Saga18} (see Section~\ref{intro}). The analysis of charge state distributions for ERs of different masses and velocities showed that the number of charge states for the non-equilibrated component $N_{\rm neq}$ exceeds those for the equilibrated one $N_{\rm eq}$ by a factor of 2--4. Mean charges $q_{m}^{\rm neq}$ and standard deviations $\sigma_{q}^{\rm neq}$ of the non-equilibrated component also exceed the same parameters of the equilibrated one $q_{m}^{\rm eq}$ and $\sigma_{q}^{\rm eq}$ within the same factor \cite{Saga08,Saga18}. In subsequent simulations for $^{252}$No ERs escaping the $^{208}$Pb target, it was assumed that $N_{\rm neq} = 2 N_{\rm eq}$. Mean charge and standard deviation values were assumed to be corresponding to $q_{m}^{\rm neq} = 2.5 q_{m}^{\rm eq}$ and $\sigma_{q}^{\rm neq} = 2.5 \sigma_{q}^{\rm eq}$.

\begin{figure}[h!]
\vspace{0.5mm}
\centerline{\includegraphics[width=0.45\textwidth]{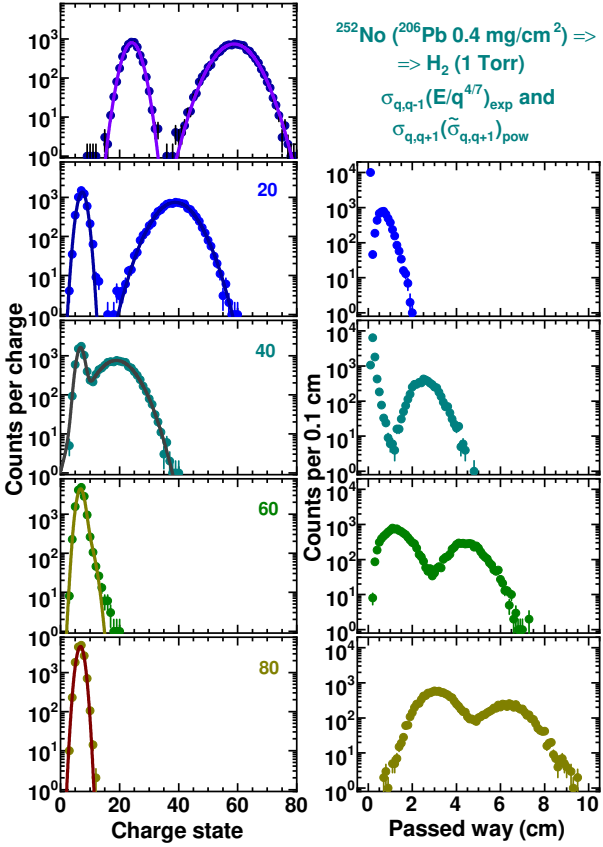}}
\vspace{-2.5mm} \caption{\label{Nochardist}The evolution of the charge distribution for the ionized $^{252}$No ERs knocked out from the 0.4 mg/cm$^{2}$ $^{206}$Pb target as they move through the H$_{2}$ gas (left panels). It starts from the initial one assumed for the ERs (upper left panel) and evolves with the increasing number of charge-changing collisions (indicated in the left panels) according to the charge-changing cross-sections shown in Fig.~\ref{252NoCS} (upper panel). The distributions of passed ways from the target for the escaped $^{252}$No ERs, which correspond to the respective number of collisions indicated in left panels, are shown in right panels.}
\end{figure}

In Fig.~\ref{Nochardist}, the evolution of the charge distribution for the ionized $^{252}$No ERs knocked out from the 0.4 mg/cm$^{2}$ $^{206}$Pb target into the H$_{2}$ gas under the pressure of 1 Torr is shown. It starts from the assumed initial distribution (upper left panel in the figure) obtained taking into account the energy distribution for ERs (Fig.~\ref{EdisNo}). Its evolution is further considered in a similar way as was done above for fixed input charges, using charge-changing cross-sections shown in Fig.~\ref{252NoCS} (upper panel). In Fig.~\ref{Nochardist}, charge distributions (left panels) and distributions of passed ways from the target (right panels), are shown for the definite number of the charge-changing collisions occurred with No ERs. The distributions of the passed ways in the charge-changing regime are close to the similar ones for distances from the target, taking into account a preferred movement of ERs in forward direction (see Fig.~\ref{AndiNo}). Note that the energy losses of No ERs along the passed ways obtained in simulations are negligible at the respective H$_{2}$ pressure. As one can see, after 60 collisions, the charge distribution becomes close to the equilibrated one with the mean charge approaching the value obtained above (Fig.~\ref{Nofixchar}). The number of collisions is a little more than the one obtained at $q_{\rm inp} = +50$ due to the presence of higher charge states in the non-equilibrated component of the input distribution. Its presence also leads to the two-humped distribution of passed ways for ERs with fully-equilibrated charges achieved after 80 collisions (Fig.~\ref{Nochardist}).

\begin{figure}[h!]
\vspace{0.5mm}
\centerline{\includegraphics[width=0.45\textwidth]{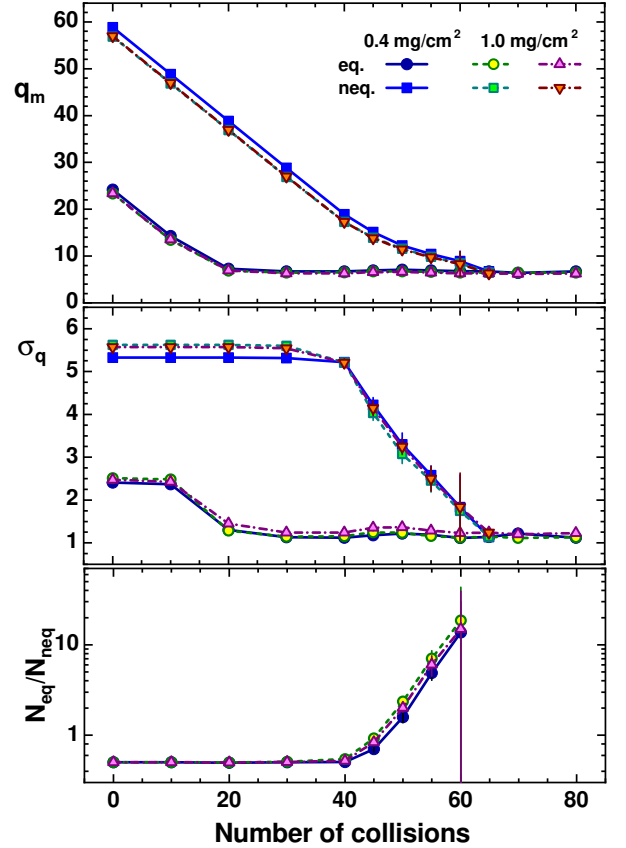}}
\vspace{-2.5mm} \caption{\label{fitpar}Parameter values of double-Gaussian fits to the charge distributions obtained in simulations for the ionized $^{252}$No ERs knocked out from the $^{206}$Pb target of the thickness of 0.4 and 1.0 mg/cm$^{2}$ (filled and open symbols, respectively) are shown as the respective functions of the number of collisions. Results of simulations obtained with the charge-changing cross-sections shown in Fig.~\ref{252NoCS} (upper panel) are plotted by circles and squares for the equilibrated (eq.) and non-equilibrated (neq.) component, respectively. Similar results obtained with the cross-sections shown in Fig.~\ref{252NoCS} (bottom panel) are plotted by triangles.}
\end{figure}

A quantitative analysis of thus obtained charge distributions for different charge-changing cross-sections and target thicknesses was performed with double-Gaussian fits to the simulated data. The examples of the results of such fits are shown in Fig.~\ref{Nochardist}. The values of fitted parameters as functions of the number of collisions $n_{\rm col}$ leading to the charge-state changes are shown in Fig.~\ref{fitpar}. As one can see in the figure, charge equilibrium in gas for the initial distribution corresponding to the `solid' equilibrated charge component is achieved at $n_{\rm col} \gtrsim 20$. The means of $q_{m}$ and $\sigma_{q}$ become such as those inherent in charge equilibration in hydrogen gas. A similar equilibrium for the initial charge distribution corresponding to the non-equilibrated component occurs at $n_{\rm col} \gtrsim 60$. According to this consideration, the non-equilibrated component in the charge distribution is still resolved at $n_{\rm col} = 60$ with $N_{\rm eq}/N_{\rm neq} \simeq 14$. But it cannot be extracted from the simulations corresponding to $n_{\rm col} \gtrsim 65$, since the only equilibrated component is manifested in these simulations.

The effect of the target thickness on the charge equilibration of ERs is tested with the same charge-changing cross-sections, as in the previous case. Increasing target thickness is important not only for a new DGFRS \cite{Popeko16,Beeckman19}, as mentioned in Section~\ref{intro}, but for any GFRS bearing in mind the future experiments on the synthesis of SHN with $Z > 118$. Increasing target thickness is one of the ways to rise the production yield of SHN, which can be synthesized in fusion-evaporation reactions with sub-pb cross-sections, as estimated in theory (see, for example, \cite{ZagGrie15,SanSaf17} and Refs. therein). In Fig.~\ref{fitpar}, the change in charge distribution parameter values for the ionized $^{252}$No ERs knocked out from the 1.0 mg/cm$^{2}$ $^{206}$Pb target is shown. Simulations were performed in the same way as the previous ones, but using the ER energy distribution obtained for the 1 mg/cm$^{2}$ target (see Fig.~\ref{EdisNo}). The ER energy distribution has an impact on the initial charge distribution. Very similar parameter values were obtained in this case, as compared with the previous ones corresponding to the 0.4 mg/cm$^{2}$ target. Similar distributions of passed ways were also obtained. For example, after 80 collisions, all $^{252}$No ERs with the equilibrated charges occupy the same 0--10 cm distance from the target (see the right bottom panel in Fig.~\ref{Nochardist}). One can state that the difference in the target thickness has a minor effect on the charge equilibration of the ionized ERs in the rarefied H$_{2}$ gas.

According to Fig.~\ref{252NoCS}, electron capture cross-sections for No ions with $q \gtrsim +20$, which correspond to the approximation with the scaling \cite{Knud81}, are about three times higher than those, corresponding to the similar one presented in Fig.~\ref{CScapQQm}. At the same time, electron loss cross-sections for the same $q$ values, which correspond to the approximation of the reduced values \cite{Franz81}, are about three times lower than the similar one corresponding to the $\sigma_{q,q+1}/\sigma_{q,q-1}$ ratio presented in Fig.~\ref{CSlocapQQm}. In this connection, the effect of the difference in the cross-section values on the equilibration of $^{252}$No charge distributions and corresponding passed ways was also tested with the pair of cross-sections shown in the bottom panel of Fig.~\ref{252NoCS}. The evolution of the charge distribution for the ionized $^{252}$No ERs knocked out from the 1.0 mg/cm$^{2}$ $^{206}$Pb target was considered in the same manner as in previous cases, but with this pair of cross-sections. The results of the analysis of charge distributions obtained in these simulations are also shown in Fig.~\ref{fitpar} by open triangles. As one can see, no significant differences in the parameter values were obtained. One can state that difference in the charge-changing cross-sections for $q \gtrsim +20$ has a minor effect on the charge equilibration of ionized ERs in the rarefied H$_{2}$ gas. At the same time, the distribution of passed ways obtained after 80 collisions for fully equilibrated $^{252}$No ions extends to 12 cm, as the result of the relatively low electron capture cross-sections for large $q$.

Concluding this section, one can state that the charge-changing cross-sections and target thickness lead to minor changes in the charge distributions resulted in the simulation of the equilibration process started from the initial charge-state distribution assumed for heavy ERs. In all cases, after 65 charge-changing collisions, the charge distribution becomes close to the equilibrated one, occupying the distance of 0--12 cm from a target. The mean charge is approaching the value determined by the equality of $\sigma_{q,q+1}=\sigma_{q,q-1}$. A rather large number of collisions is caused by the equilibration process of the non-equilibrated component present in the initial charge state distribution. High capture cross-sections for high charge states of the non-equilibrated component cause rather small passed ways from the target for escaped ERs. At the same time, the mean value of the passed ways in gas, which corresponds to the re-charging of the charge-equilibrated ERs varies from 0.08 to 0.12 cm with about the same standard deviations (the values depend on the charge-changing cross-sections used in simulations). These circumstances allow one to use the equilibrated charge state distribution in trajectory simulations for heavy ERs passing through magnetic elements filled with rarefied H$_{2}$ gas.

\section{\label{Summa}Summary}

In the present work, the charge-changing process in rarefied H$_{2}$ gas, which leads to the charge state equilibration of ionized heavy evaporation residues (ERs) produced in the fusion-evaporation reactions is considered. It is assumed that ERs knocked out from a target by heavy ion (HI) projectiles have a very broad charge state distribution. Besides the equilibrated `solid' component \cite{SIM82,SY94,SG01}, ERs have much higher charge states corresponding to the non-equilibrated component \cite{Steig63,Scob83,Brink94,Saga08,Saga18}. This component arises from the inner atomic shells ionization induced by the conversion of nuclear transitions in ERs. Vacancies formed in the inner shells lead to the Auger cascades, which significantly increase the ion charges of ERs over the equilibrated ones. Transformation of this component into the equilibrated one in rarefied H$_{2}$ may occur at large distances from the target due to the number of electron captures leading to charge equilibration.

Mean equilibrated charges of HIs in rarefied H$_{2}$ gas were estimated using two empirical formulae derived in the present work with the fits to available experimental data. The first one was the result of the joint fit to the HI data \cite{WitBetzAD73,Schar17} and the data obtained in experiments with the Dubna gas-filled recoil separator (DGFRS) for heavy ERs \cite{DGFRScha}. These equilibrated charges corresponded to the $q_{m}(\rm fit)$ values. The second formula was derived with the DGFRS data only \cite{DGFRScha} relating to the $q_{m}^{\rm DGFRS}$ equilibrated charges. A comparison of the $q_{m}$ values obtained with these formulae showed their mutual agreement within $\pm$10\% at the velocities of $1.8 \leqslant v/v_{0} \leqslant 3.0$.

A way to charge equilibration is determined by single-electron capture $\sigma_{q,q-1}$, and loss $\sigma_{q,q+1}$ cross-sections, which are changed with the charge state of ERs moving in gas. Several (semi-)empirical approaches to the estimates of charge-changing cross-sections \cite{Knud81,Schla83,Cornel06,Franz81} were examined with the aim to choose the best one describing the HI data at the energies of 0.035--1.4 MeV/nucleon \cite{Betz73,Soren84,Hvelp92,Franz81,Berk81}. The exponential function fitting the $\sigma_{q,q-1}/q$ data against the $E/q^{4/7}$ values (scaling proposed in \cite{Knud81}) was making the best among others according to $\chi^{2}_{r}$ criteria. For the electron loss cross-sections, an empirical formula proposed by Franzke \cite{Franz81} revealed a rather good scaling of low energy data using the $q/q_{m}(\rm fit)$ values as an argument. Reduced electron loss cross-sections $\tilde{\sigma}_{q,q+1}$ presented as a function of $q/q_{m}(\rm fit)$ were well fitted with a power function.

New empirical formulae for the $\sigma_{q,q-1}$ and $\sigma_{q,q+1}$ cross-sections were also proposed, which disclosed a rather good agreement with experimental data according to the $\chi^{2}_{r}$ criteria. A proper scaling was thus obtained for the $\sigma_{q,q-1}$ data in the dependence on $x = q/q_{m}^{\rm DGFRS}$. The available data were well fitted with the exponential function in the form of $a\exp[b/(x + c)]$. A quite reasonable scaling was also obtained for the $\sigma_{q,q+1}/\sigma_{q,q-1}$ ratios displayed as a function of $q/q_{m}(\rm fit)$. The ratios were well fitted with a power function.

In the application of the obtained formulae to the ionized $^{252}$No ERs produced in the $^{206}$Pb($^{48}$Ca,2$n$) reaction, the approximations proposed for the $\sigma_{q,q-1}(q)$ cross-section showed a difference corresponding to a factor of 3 for the calculated values at $q \gtrsim +20$. A similar mutual disagreement was also obtained for the $\sigma_{q,q+1}(q)$ cross-section approximations. For the $q \ll q_{m}$ charge states, $\sigma_{q,q+1}(q)$ values thus obtained differed from each other even more. This difference is not critical in the initial stage of the equilibration process for highly ionized ERs.

Monte Carlo simulations, similar to those used earlier \cite{Ninov95,Paul89,Gregor13}, but based on the approximations obtained for charge-changing cross-sections, allowed one to get an idea of the rapidity of the equilibration process for initially highly ionized ERs. These simulations showed, by the example of $^{252}$No, that the `solid' equilibrated charge component with $q_{m}\simeq$ 24 becomes the equilibrated one in rarefied H$_{2}$ gas with $q_{m}\simeq$ 6 after $\sim$30 charge-changing collisions corresponding to the passed way of 2--5 cm from the target. Much higher charge states of the non-equilibrated  component become the equilibrated one in gas after $\sim$65 collisions corresponding to the same passed way from the target. The charge equilibration of ionized heavy ERs poorly depends on the target thickness. Such fast (short-range) equilibration allows one to use mean equilibrated charges in simulations of ERs transmission through gas-filled magnetic systems.

A similar approach could be applied to the consideration of charge equilibration for ionized ERs in rarefied He gas, bearing in mind a number of gas-filled separators working with He \cite{Miyat87,Ghio88,Ninov95,Leino95,TASCA08,SHANS13,Gregor13}. In this case, distances between successive charge-changing collisions for HIs do not differ so much from those estimated for H$_{2}$. Indeed, an exponential approximation to the $\sigma_{q,q-1}$ data \cite{Betz73,Knud81,Datz90} scaled according to \cite{Knud81} gives us cross-section values in the region of (3.5--160)$\cdot$10$^{-16}$ cm$^{2}$ for the initial $^{252}$No charge-states in the range between +10 and +80. These cross-sections lead to the respective distances between charge-changing collisions in the range of (8.7--0.19)$\cdot$10$^{-2}$ cm for the He atomic density of 3.3$\cdot$10$^{16}$ cm$^{-3}$ under the pressure of $\sim$1 Torr. Data on $\sigma_{q,q+1}$ for HIs passed through He should be considered and approximated, bearing in mind their importance in the cross-section estimates for the charge states in the vicinity of the ER equilibrated charge.

\begin{acknowledgments}
The author is indebted to Dr. V.K. Utyonkov for providing new data on the mean charges for the heaviest ERs in hydrogen, which were obtained with DGFRS. This study was supported in part by the directorate of JINR under a special grant for the SHN program.
\end{acknowledgments}

\end{document}